\documentstyle[11pt,aaspp]{article}

\slugcomment{Accepted by the {\it Astronomical Journal}}

\def\kms{\hbox{$\rm {km}~\rm s^{-1}$}}
\def\ha{\hbox{H$\alpha$}}
\def\hi{\ion{H}{1}}
\def\hii{\ion{H}{2}}
\def\cmsqs{\hbox{$\rm{cm}^{-2}~\rm{s}^{-1}$}}
\def\ergcmsqs{\hbox{$\rm{erg}~\rm{cm}^{-2}~\rm{s}^{-1}$}}
\def\cmsq{\hbox{$\rm{cm}^{-2}$}}
\def\cmcub{\hbox{$\rm{cm}^{-3}$}}
\def\cmsqssr{\hbox{$\rm{cm}^{-2}~\rm{s}^{-1}~\rm{sr}^{-1}$}}
\def\cmsixpc{\hbox{cm$^{-6}$~pc}}

\def\msun{\hbox{$M_{\odot}$}}
\def\rahr{\hbox{$^{\rm h}$}}
\def\ramin{\hbox{$^{\rm m}$}}

\def\degk{\hbox{$^{\circ}{\rm K}$}}
\def\ni{\noindent}
\def\etal{{\it et al.\ }}
\def\eg{{\it e.g.\ }}

\begin{document}

\title{Detection of H$\alpha$ Emission from the Magellanic Stream: \\
Evidence for an Extended Gaseous Galactic Halo}

\author{Benjamin J. Weiner\altaffilmark{1} and T. B. Williams\altaffilmark{1}}

\affil{Department of Physics and Astronomy, Rutgers, The State University \\
Box 849, Piscataway, NJ 08855-0849 \\
bweiner@physics.rutgers.edu, williams@physics.rutgers.edu}

\altaffiltext{1}{Visiting Astronomer, Cerro Tololo Inter-American Observatory.
CTIO is operated by AURA, Inc.\ under contract to the National Science
Foundation.}

\quad\newline
\centerline{Rutgers Astrophysics Preprint No. 180}

\begin{abstract}
We have detected faint \ha\ emission from several points along the
Magellanic Stream, using the Rutgers Fabry--Perot Interferometer at
the CTIO 1.5-m telescope.  The sources of the emission are diffuse;
at each observed position, there is no variation in intensity over the
7\arcmin\ field of the Fabry--Perot.  At points on the leading edges
of the \hi\ clouds MS II, MS III, and MS IV, we detect
\ha\ emission of surface brightness
$0.37 \pm 0.02$~Rayleighs, $0.21 \pm 0.04$~R, and $0.20 \pm 0.02$~R
respectively, corresponding to emission measures of 1.0 to 0.5~\cmsixpc.
We have observed several positions near the MS IV concentration, and
find that the strongest emission is on the sharp leading-edge density
gradient.  There is less emission at points away from the gradient,
and halfway between MS III and MS IV the \ha\ surface brightness is
$< 0.04$~R.

We attribute the \ha\ emission at cloud leading edges
to heating of the Stream clouds by ram pressure from ionized gas
in the halo of the Galaxy.  These observations suggest that ram
pressure from halo gas plays a large role in stripping the Stream out
of the Magellanic Clouds.  They also suggest the presence of a
relatively large density of gas, $n_{\rm H} \sim 10^{-4}~\cmcub$,
in the Galactic halo at $\sim 50$~kpc radius, and far above the
Galactic plane, $\vert b\vert \sim 80\deg$.  This implies that the
Galaxy has a very large baryonic, gaseous extent, and supports models
of Lyman-$\alpha$ and metal-line QSO absorption lines in which the
absorption systems reside in extended galactic halos.

\end{abstract}

\keywords{galaxies: Magellanic Clouds -- Galaxy: corona of -- Galaxy: halo of
-- quasars: absorption lines}

\section{Introduction}

The Magellanic Stream is a long filament of \hi\ clouds which stretches
over 100\deg\ on the sky,
trailing behind the Magellanic Clouds in their orbit around the
Galaxy (Mathewson \etal 1974).
Observations of the Stream at 21 cm wavelengths show that it is a chain
of clouds, connected by lower--density gas; the clouds have been labeled
MS~I through MS~VI by Mathewson \etal (1977).  These clouds
generally have a high-density concentration with
a relatively sharp density gradient on the leading edge, where the
leading edge is determined by the direction of the proper motion
of the LMC (Jones \etal 1994).  The Stream has no known stellar component
(Recillas-Cruz 1982, Br\"uck \& Hawkins 1983, Mathewson \etal 1979);
it has previously been detected only in the
21 cm \hi\ line, and in absorption against a background galaxy (Songaila 1981,
Lu \etal 1994).  Hence, its distance
is unknown, although its leading end, MS I, connects to the Magellanic Clouds
and is presumably at 50--60 kpc; recent estimates for the distance of the
tip, MS VI, range from 20 kpc (Moore \& Davis 1994) to 60 kpc
(Gardiner \etal 1994).

There are many candidate explanations for the origin of the Magellanic
Stream; most invoke either tidal or ram-pressure forces to detach the
Stream from the Clouds.  In tidal models, the Stream
is torn out of the Magellanic Clouds by gravitational tides variously
attributed to: the Galaxy ({\it e.g.}
Lin \& Lynden-Bell 1977, Gardiner \etal 1994, Lin \etal 1995), an
encounter between the LMC and SMC
(Murai \& Fujimoto 1980), or an encounter with M31 (Shuter 1992).
A weakness of tidal models is that stars should also be affected by tides,
yet no stars appear to be associated with the Stream.

In ram-pressure models, the Stream is swept out of the Magellanic Clouds
by gas postulated to exist in the Galactic halo, such as a diffuse
ionized corona ({\it e.g.} Bregman 1979, Meurer \etal 1985, Sofue 1994),
or an extended ionized disk and halo (Moore \& Davis 1994).
These models have had to invoke halo gas {\it ad hoc}.
Another difficulty is that the timescale for a stripping instability
to develop may be very long for reasonable halo gas densities (Bregman 1979).
It is clear that these
scenarios for the formation of the Stream are highly dependent on
unknowns such as the mass and extent of the Galactic halo and the
putative ionized Galactic corona.  Conversely, the Stream can
be a probe of these unknowns.

We report here the results of high-sensitivity observations of \ha\
emission from the Magellanic Stream, using the Rutgers Imaging
Fabry--Perot interferometer (RFP).  The Fabry--Perot simultaneously
provides spectral coverage over a short wavelength interval, and a
large collecting area, making it well suited to search for faint diffuse
emission lines (Williams 1994).

In section 2 we describe the observations and data reduction.  In
section 3 we summarize the results and \ha\ detections, and show that
the emission is associated with cloud leading edges.  In section 4, we
consider possible sources for the emission and show that the most
plausible source is heating of the Stream gas by ram pressure,
arising as the clouds of the
Stream collide with diffuse gas in the Galactic halo.  This diffuse
gas is probably at the virial temperature of the halo,
$T \sim 2 \times 10^6~\degk$, implying the existence of a tenuous, hot
ionized Galactic corona.

\section{Observations}

We used the RFP at the f/7.5 Cassegrain focus of the CTIO 1.5-m
telescope.  The etalon used has
resolution $\sigma = 15~\kms$, 0.75~\AA\ FWHM at \ha, and the
free spectral range is 22~\AA.  The detector was the Tek 1024 \#2 CCD, binned
$2 \times 2$, giving 1.94\arcsec\ pixels; the gain was 1.7 electrons/ADU
and the read noise was 4.9 electrons.  The field of view of the RFP is
7\arcmin\ in diameter; there is a 5.5 \AA\ gradient in wavelength
from center to edge of the field.  We used a blocking filter of
central wavelength 6560~\AA\ and FWHM 12~\AA, which ensures that
only one Fabry--Perot order is imaged.  The images are flat-fielded
using IRAF.\footnote{IRAF is distributed by NOAO, which is operated by
AURA Inc., under contract to the NSF.}

The observations were made on the nights of 12--16 August, 1994.
Table 1 lists the object fields observed and summarizes the results.
The individual exposures
were 15 minutes each; in order to achieve accurate sky subtraction, we
took exposures of sky fields 8\deg--15\deg\ away from each object,
chosen to be in regions free of high-velocity \hi.  Generally we
sequenced exposures so that each object
spectrum has its corresponding sky exposure obtained immediately before
or after.  Individual exposures
on the same field were offset in different directions by 30\arcsec\
to avoid any spatial structure which might mimic an emission line.
Exposures of a hydrogen lamp were taken every hour for wavelength
calibration.  The weather was excellent; all five nights were
photometric.  The flux calibration was derived by observing the
planetary nebula NGC~6302 (Acker \etal 1993), scanning the
\ha\ emission line.

Since the center-to-edge gradient is considerably larger than the
resolution of the etalon, any diffuse emission line which fills
the field will appear as a ring around the optical axis of the etalon.
We divide the image into circular annuli with width corresponding to
0.1~\AA, and estimate the flux within each annulus, to
obtain a 5~\AA\ long section of the spectrum of the field.
Each annulus contains the same number of pixels, due
to the parabolic variation of wavelength with radius.

We estimate the average flux
per pixel in each annulus using the biweight statistic (Beers \etal
1990); the biweight gives very little weight to outliers, which
effectively clips stars and cosmic rays.  We
derive an error estimate from the
biweight scale, a robust analog of the standard deviation.
This estimated error, derived directly from the dispersion
among the pixels in each sample, is consistent with that expected from
the read noise and photon statistics.

We pair each object-field spectrum with a sky-field spectrum and
subtract to isolate the object contribution.  There is generally an
offset in the continuum due to temporal variations, differences in
airmass, and scattered moonlight, so we subtract a constant continuum
level, determined by taking the biweight of a region of the spectrum
away from any object or atmospheric emission lines.  Figure 1 shows an
example: the spectra from one pair of object and sky exposures for the
MS II A field, and the residual object spectrum, with the
signature of an emission line.  The observed wavelength agrees with
the LSR velocity of the \hi\ observed by Morras (1985).
The strong feature at 6563~\AA\ is the geocoronal \ha\ line,
demonstrating the sensitivity of the Fabry-Perot, since this is a rather
weak atmospheric feature (Osterbrock \& Martel 1992).  The emission from
MS II is visible at 6560~\AA.  The height of the peak corresponds to 2.5
electrons per pixel; the ring signature of the MS II emission line is actually
visible in the raw object-field CCD frame.  The intensity of the \ha\ line
from MS II A is 0.37 Rayleighs (1 Rayleigh = $10^6~{\rm photons}~\cmsqssr$),
or an emission measure (EM) of 1.0~\cmsixpc, if the gas is at $10^4~\degk$.
The total combined spectrum of MS II A, shown in Figure 2(a), is
composed of six object--sky pairs like that shown in Figure 1(c).

\section{Results}

For each object field observed, we combined the object-minus-sky
pairs, first subtracting a continuum from each pair.  In Figure 2(a-c),
we present the total combined spectra for fields MS II A, MS III, and
MS IV C, which are located at the leading edges of \hi\ density
concentrations within each cloud.
We also plot a LOWESS fit to the data; LOWESS
(locally weighted scatterplot smoothing) is a robust smoothing
algorithm (Cleveland \& McGill 1984).
In each of the three fields, a \ha\ emission line is visible, at a velocity
which agrees with the \hi\ velocity from observations at 21 cm.
The intensities, derived from fitting Voigt profiles to the spectra,
are $0.37 \pm 0.02$~Rayleighs, $0.21 \pm 0.04$~R, and $0.20 \pm 0.02$~R
from MS II A, MS III, and MS IV C respectively.
The features around 6563~\AA\ in the MS II A spectrum are due to
incomplete cancellation of the geocoronal \ha\ line, which shows
large temporal variations.

The high negative velocity of MS VI, combined with an unfavorable
LSR velocity correction, places it on the wing of the OH 6553.6~\AA\
line; we can only set an upper limit of 0.4 Rayleighs on emission from
this field, and will exclude it from further discussion.
We also observed a field, MS II B, located about midway
between the MS II A and MS III fields, which is on the Stream
but not on a cloud leading edge.  This spectrum is shown in Figure 3(a);
the ``bump'' at 6560.25~\AA\ is only marginally significant,
having intensity $0.07 \pm 0.02$~R, which suggests
that the \ha\ emission is strongest on the cloud leading edges.

Our observations around the MS IV concentration allow us to
investigate this possibility in more detail.
The map of \hi\ surface density in this region (Figure 1 of Cohen 1982)
shows a strong wedge-shaped density gradient at 23\rahr 42\ramin\,
--12\deg, and trailing, fragmented lower-density contours to the
northwest, reminiscent of a bow shock or ram pressure stripping.
We observed four fields
along the Stream, designated A through D.  Fields A \& B are ahead
of the density gradient, C is on the gradient, and D is behind it.
The spectra of A, B, \& D are presented in Figure 3(b-d), and the
spectrum of C is in Figure 2(c).  For
these fields, time constraints forced us to use the same set of four sky
exposures for all three object fields.  The sky subtraction is poorer,
leading to an artificial ``dropoff'' at both ends of the spectrum.

However, this does not affect the conclusion we draw, which is that
the emission clearly is weak at points not on the density
gradient.  Fields A \& B have no significant signature of emission,
with $2\sigma$ upper limits of 0.04~R and 0.06~R.  Field D shows marginal
evidence for a double-peaked profile; fitting two Voigt profiles to
the peaks yields an intensity of $0.09 \pm 0.03$~R.
The emission from the MS IV C field is much stronger (Figure 2(c)),
confirming that the emission is associated with the cloud leading edges.
The line profile of the emission from MS IV C is broadened and
fairly asymmetric, suggesting that the line of sight passes through at
least two components at different velocities.

The results of our observations, including \ha\ velocities and intensities,
are summarized in Table 1.  There are firm detections of \ha\ emission
from MS II A, MS III, and MS IV C, with intensity
$0.37 \pm 0.02$~R, $0.21 \pm 0.04$~R, and $0.20 \pm 0.02$~R respectively.
The fluxes are 2.6, 1.4, and
1.4~$\times~10^{-17}$~erg cm$^{-2}$ sec$^{-1}$ arcsec$^{-2}$.
If the Stream gas is at $\sim 10^4~\degk$,
the emission measures (EM) are 1.0, 0.5, and 0.5 \cmsixpc.
The EMs imply ionization fractions of $x \sim 0.5$ in the leading edges.
On the leading edges, the \ha\ velocities are in good agreement with
the \hi\ velocities, and the velocity dispersions are 15 -- 30 \kms.
There is at best marginal evidence for \ha\ emission on the points
not on the leading edges.
(These leading edges are defined by the \hi\ maps for MS II, III, and IV,
taken from Morras (1985), Mirabel, Cohen \& Davies (1979), and
Cohen (1982) respectively.)

The \ha\ intensity, velocity, and velocity dispersion are derived from a
Voigt profile fitted to the combined spectrum, allowing for a gaussian
velocity dispersion convolved with the instrumental Voigt profile
of the RFP.  They may be affected by non-gaussian velocity substructure
such as
blending along the line of sight, as illustrated by the MS IV C spectrum;
substructure in the line profile has a small effect on the derived intensity,
but it can have a large effect on the derived velocity dispersion.
We searched for spatial structure in the emission by dividing the
images into quadrants and fitting a Voigt profile to each quadrant
separately; no significant variations in intensity were seen.
The formal errors in the \ha\ velocities yielded by the fit are in all
cases less than 5~\kms.

The important results from the observations are: (1) the
\ha\ emission is associated with leading edges of the clouds,
since the detections are all located on leading edges,
and there are no compelling detections on the non-leading edge
fields, and (2) the \ha\ intensity does not correlate with the
observed \hi\ column density.

For all fields in which we observe significant emission, the \ha\
velocity agrees with the \hi\ velocity (see Table 1).  The
velocities are sufficiently high that Galactic contamination is
implausible.  Since we observe emission in several different fields at
different wavelengths, we are certain that the emission cannot be a
weak atmospheric feature.  The sky-subtraction process is relatively
simple, and we have taken care to alternate object-field and sky-field
exposures, to avoid temporal variations.  We conclude that the
observed signatures are genuine \ha\ emission from the Magellanic
Stream.

\section{Sources of Emission -- Contact with coronal gas}

The similarity of the \ha\ intensity on the leading edges of MS II, III,
and IV motivates us to seek a common cause.
We will show that the most likely cause is emission generated as
the Stream clouds are heated by kinetic and thermal energy input,
presumably caused by the motion of the Stream clouds
through a galactic corona of hot ionized gas.

The \ha\ emission could result from collisional ionization caused by
some form of mechanical energy input; the most plausible source is
contact between the Stream clouds and an lower-density ambient ionized
medium.  The \ha\ emission could be powered by ram pressure heating as
the clouds move through ambient gas in the Galactic halo, or by
thermal conduction from hot, ionized gas, or both.  Notably, ram
pressure sweeping may explain the shapes of the leading edges of the
LMC and the Magellanic Stream clouds, and the associated \hi\ density
gradients, as suggested by Mathewson \etal (1977).

\subsection{Ram pressure heating}

The tapered shape of the Stream clouds and the H I density gradients on
their leading edges suggest that the clouds are moving through lower density
ambient gas.  If so, the ram pressure from the ambient gas
will transfer energy to the Stream clouds, heating their leading faces.
This would naturally explain the association of \ha\ emission with leading
edges.  However, it is difficult to calculate the efficiency of the
process in which energy input due to ram pressure is converted to energy
output by radiation.

\subsubsection{A simple drag model}

We can make a crude estimate by assuming that the Stream clouds are
subject to a drag force from the lower-density ambient gas.
As a simple model, we assume that the cloud does not accrete
material from the hot medium, and that the kinetic energy
the cloud loses to drag goes into heating the leading face
of the cloud.

Approximating the cloud as a solid body moving through the hot
gas, we express the drag in the standard form:

\begin{equation}
\frac{dp}{dt} = F_{drag} = - {\textstyle{1\over2}} C_D \rho_1 A v^2,
\end{equation}

\ni where $C_D$ is a dimensionless drag coefficient dependent on the
cloud shape, $\rho_1= 1.4 m_{\rm H} n_{\rm H}$ is the density of
the coronal gas, $A$ is the frontal area of the cloud, and $v = 220~\kms$
is its velocity relative to the corona.  If the coronal gas is
at the virial temperature of the halo, the Stream clouds are moving
through it at approximately the sound speed in the gas.  In this
transonic regime, the drag is relatively high, and there is not a
stand-off shock ahead of the cloud; rather, we expect the coronal
gas to transfer momentum to the Stream cloud at or close to its face.
Thus the kinetic energy lost by the cloud is likely to be dissipated
into its leading face.  Taking $C_D = 1$ and
assuming the energy is distributed uniformly over the cloud face,
the input energy flux $\Sigma_{in}$ is

\begin{equation}
\Sigma_{in} = {\textstyle{1\over2}} \rho_1 v^3.
\end{equation}

We now assume that this energy is all converted to radiation, and
consider the case of the MS IV leading face.  The flux
in \ha\ from the MS IV C field is
$\Sigma_{obs} = 6.0 \times 10^{-7}$~\ergcmsqs.  MS IV has a opening
angle of $\sim 90\deg$, so we assume that the emission is enhanced
by a factor of 3 due to the depth of the heating zone along our line of
sight.  Then the ambient gas density required to supply the energy emitted
in the \ha\ line {\it alone} ($\Sigma_{in} = \Sigma_{obs}/3$) is
$n_{\rm H} (\ha) = 1.5 \times 10^{-5}~\cmcub$.

If the Stream gas is at $T \sim 10^4$ K, as suggested by its velocity
width and the presence of \ha, there are 2.2 recombinations
per \ha\ photon (Pengelly 1964; Martin 1988).  Since an \ha\ photon
carries off only one-seventh of the energy of a recombination, and
energy must also go into metal emission lines and internal motions in
the cloud, the required density of diffuse gas in the halo is on the
order of $n_{\rm H} \sim 10^{-4}~\cmcub$ or greater.

This model involves a chain of assumptions: that the coronal
gas exerts drag on the cloud with $C_D \sim 1$; that the drag heats the
face of the cloud; and that this heat will be efficiently converted
into radiation.  The first assumption is justified for a
blunt object such as a gas cloud, in the low viscosity
regime applicable here (\eg Figure 3.15 of Tritton 1988).
The drag acts by increasing the pressure on the leading face
of the cloud (\eg Figure 12.9 of Tritton 1988), so that it
is reasonable to assume that the kinetic energy lost to drag
by the cloud is converted to heat at the face.\footnote{Follow-up
observations in August 1995, to be reported
fully in a later paper, confirm that \ha\ emission of 0.05 -- 0.2~R
is distributed over the leading face of the MS IV cloud.}
Finally, the \ha\ and H I velocity dispersions indicate
that the cloud gas is at $T \sim 10^4~\degk$, so it will quickly lose
the heat energy to radiation, due to the steepness of the cooling
curve above $10^4~\degk$.

The model demonstrates that the drag from coronal gas of density
$n_{\rm H} \sim 10^{-4}~\cmcub$ can generate an energy input to the
MS cloud that is of the right order of magnitude to power the
observed \ha\ emission.  This is only an order of magnitude estimate,
as there are a number of competing factors whose values are uncertain,
such as the drag coefficient and the efficiency of conversion of
kinetic energy to radiation.  Nonetheless, it is significant
that the implied coronal density is similar to
that inferred by Wang (1992) from observations of the X-ray background.

\subsubsection{Instability of the flow}

The drag analysis above treats the Stream cloud as a solid object
moving through coronal gas.  As a first approximation, this is
reasonable, since the Stream gas is much denser than the putative
corona.  However, it is incomplete, since the interface
between the cloud and the coronal gas may be unstable.
Even if the cloud and the corona are in pressure equilibrium,
the flow of the coronal gas past the surface of the cloud
may be subject to the Kelvin-Helmholtz instability, which
arises at the interface of two fluids with a relative shear
velocity.

The Kelvin-Helmholtz instability
occurs for all wavelengths $\lambda$ satisfying

\begin{equation}
\lambda < \lambda_{crit} = \frac{2\pi n_1 n_2 v^2}{g({n_2}^2 - {n_1}^2)},
\end{equation}

\ni where $g = G M_c / {R_c}^2$ is the gravitational acceleration at the
surface of the Stream cloud, $n_1$ and $n_2$ are the number densities of
the hot gas and the cloud, and $v$ is their relative velocity (Chandrasekhar
1961).  These formulae apply to incompressible fluids; the incompressible
case is a good approximation in all but strongly supersonic motion
(Bradshaw 1977).
The $e$-folding timescale for the growth of a perturbation is
$\tau$ (Chandrasekhar 1961):

\begin{equation}
\tau = {\left[ \frac{4\pi^2 n_1 n_2 v^2}{\lambda^2(n_1 + n_2)^2}
             - \frac{2\pi g (n_2 - n_1)}{\lambda(n_1 + n_2)}
              \right]}^{-1/2}.
\end{equation}

For the dense cloud at the head of MS IV, we assume its distance to be
50 kpc, obtaining approximate values of
$M_c = 2 \times 10^7~\msun$, $R_c = 700$~pc, and $n_{2\rm H} = 0.05~\cmcub$
(Cohen 1982), and assume the cloud velocity is $v = 220~\kms$ and the
coronal density $n_{1\rm H} = 10^{-4}~\cmcub$.  This yields a
critical wavelength of $\lambda_{crit} = 3.5$~kpc.  Since this is
larger than the size of the dense cloud, and comparable to the
width of the entire low-density component of the Stream, the
cloud-corona interface is unstable on all relevant length
scales.\footnote{The instability can be suppressed by a magnetic field
of strength $B_\parallel > 1~\mu\rm{G}$ everywhere parallel to the
shear velocity, but this is unlikely in gas of such low density.}

Representative timescales for the instability to develop are
$\tau_{10} = 1.6 \times 10^5$~yr and $\tau_{1000} = 1.9 \times 10^7$~yr
for perturbations of wavelength 10 pc and 1 kpc respectively.
These are shorter than the orbital timescale,
$\tau_{dyn} = {\rm several} \times 10^8$~yr, and therefore
the interface between the Stream cloud and the coronal gas will
become complex and turbulent.  This increases the likelihood
that coronal gas can transfer energy to the cloud; it also
suggests that the cloud will itself be ram pressure stripped
by the corona.  The \hi\ observations of the Stream show
dense clouds trailed by diffuse \hi\ with a complex distribution
(\eg Figure 1 of Cohen 1982), suggesting that the clouds are
indeed being stripped by ram pressure.  Since the cloud sizes
are a few kpc, the stripping will occur over timescales
$\sim \tau_{1000}$.  This may contribute to the observed
decrease in peak \hi\ column density along the Stream from MS~I to MS~VI.

\subsection{Thermal conduction}

If the clouds of the Magellanic Stream are surrounded by coronal
ambient gas, it is presumably ionized and quite hot, at the
virial temperature of the Galactic halo.  Thermal conduction
at the cloud--corona interface will heat the cloud and produce an
ionized zone at the surface, in which recombination and \ha\ emission
can occur.

If the clouds of the Magellanic Stream are in contact with a hot
corona of $T = 1.7 \times 10^6~\degk$, the virial temperature of the
Galactic halo (Fall \& Rees 1985), and the corona has density
$n_{\rm H} = 10^{-4}~\cmcub$,
motivated by Wang (1992), and the subclouds observed have
$R \sim 1~\rm{kpc}$, then the clouds will evaporate, and the evaporation
will be unsaturated (McKee \& Cowie 1977).  The emission measure expected
from a conduction front in an evaporating cloud is calculated
in \S IIIb of McKee \& Cowie (1977); ignoring photoionization
of the skin of the cloud, the EM is less than ${n_{\rm H}}^2 R$,
and if photoionization drives the ionization front, the EM is
$\sim 18{n_{\rm H}}^2 R = 2 \times 10^{-4}~\cmsixpc$.  This is
far below the observed EMs of 0.5 to 1 \cmsixpc, even if the
lines of sight pass through multiple conduction fronts, or if
the interface between the cloud and corona is augmented by
the Kelvin-Helmholtz instability described above.

Additionally, thermal conduction does not explain the association of
\ha\ with leading edges.  Radiation from an ionized cloud-corona
interface should be uniform across the clouds, or perhaps exhibit a
correlation with gas density, since recombinations could occur more
easily in regions of higher density.  However, the observations
show no trend of \ha\ with \hi\ column density.  Hence, although thermal
conduction may contribute, it is not the primary cause
of the observed \ha.

\section{Other Sources of Emission}

We now consider and rule out other potential sources of emission
which do not invoke the presence of Galactic coronal gas.

\subsection{Relic recombination}

Could the \ha\ be a relic of a
transient ionization phenomenon that occurred as the clouds of the
Stream were detached from the Magellanic Clouds?  This is implausible,
since the Stream, as observed in \hi, does join onto the Clouds.
Nevertheless, we can rule this out firmly by comparing the orbital
timescale to the recombination timescale.

Consider the case of MS IV, which trails the Magellanic Clouds by
$\sim 70\deg$ on the sky.  A simple lower limit to the time
since MS IV separated from the Clouds is the time $t_{min}$ it
takes the Clouds to move $70\deg$ of arc in their orbit.  Since MS IV
presumably retains some transverse velocity this should be an
underestimate.  The galactocentric transverse velocity of the LMC
is $215 \pm 48~\kms$ (Jones \etal 1994), and its distance is $\sim 50$~kpc,
giving $t_{min} \sim 2.8 \times 10^8$ yr.

If the ionization was a single event, the recombination
rate was probably greater in the past, so it is conservative to assume
a constant rate equal to that implied by the observations,
$\eta_{rec} = 4.4 \times 10^5 ~\cmsqs$.  From the observations of
Cohen (1982), the clump at the leading edge of MS IV has an angular radius
$\theta \gtrsim 1\deg$.This suggests that constant recombination
since a single drastic ionization should produce a mass of \hi\ in MS IV,

\begin{equation}
M_{\rm{HI}} > \pi(\theta D)^2 m_p \eta_{rec}t_{min} = 30,000~D^2 \msun ,
\end{equation}

\ni where $D$ is the distance to MS IV in kiloparsecs.

The observations of MS III and MS IV by Cohen (1982) detected a total
\hi\ mass of $M_{\rm{HI}} = 9000 D^2~\msun$, of which roughly half belongs
to MS IV, a factor of about 6 less than the supposed lower limit.  In
other words, steady recombination since a single ionization event would
produce much more \hi\ than is observed.  Comparing the product
$\eta_{rec}t_{min} = 3.8 \times 10^{21}~\cmsq$ (expected column density)
to Cohen's observed peak column density on MS IV,
$N_{\rm{HI}} = 1.3 \times 10^{20}~\cmsq$,
yields a similar conclusion.  The present-day \ha\ emission is too
strong to be a remnant of a single ionization event.

\subsection{Photoionization}

We now rule out present-day sources of photoionization.  The \hi\ column
densities in all observed fields are optically thick to ionizing
radiation, so any uniform ionizing flux should produce
ionization rates and \ha\ emission which are roughly constant from
field to field; this is not observed.  (This also excludes other uniform
sources such as cosmic rays.)  Additionally, the only plausible
sources of a uniform ionizing background are the extragalactic radiation
field, and any ionizing flux emergent from the Galactic disk.  Upper
limits on the ionizing flux from Fabry-Perot \ha\ observations show that these
sources cannot produce \ha\ flux as high as observed on the Stream
(Kutyrev \& Reynolds 1989, Songaila \etal 1989, Vogel \etal 1994).

The required photoionization rates could be produced by a small number
of OB stars, {\it e.g.} just one star producing
$10^{48}~\rm{photons~s}^{-1}$ located $\sim 100$ pc ($\sim 6\arcmin$)
away from the observed field.  However, such a star in the Stream
would produce a luminous \hii\ region, which would be easily detected
in surveys such as the \ha\ survey mentioned by Mathewson \etal
(1979).  Additionally it is implausible that such a star, with a
lifetime less than a few $\times~10^7$ years, could be associated with the
Stream, much less that several stars or groups should be fortuitously
located to produce similar \ha\ intensities on MS II, III, and IV.  This
coincidence of intensities would be required for any discrete sources
of photoionization.  Hence photoionization is ruled out as a cause of
the \ha\ emission.

\section{Conclusions}

The \ha\ emission from the leading edges of MS II, MS III, and MS IV
is best explained by ram pressure heating from
a surrounding medium of low-density gas; although there are several
unknown factors in this process, it is the only candidate which can
produce the right order of magnitude of energy output.

This gas is likely
to be at the virial temperature of the Galactic halo,
$\sim 1.7 \times 10^6~\degk$, and nearly completely ionized,
undetectable except in X-rays.  The similarity of the \ha\
fluxes from MS II, III, and IV shows that this ambient medium is
distributed over large scales.  This
suggests that the Stream is moving through a Galactic corona of hot,
ionized gas of density $n_{\rm H} \sim 10^{-4}~\cmcub$.

How far does this corona extend?  Although the distance of the Stream is
unknown, MS I is joined to the \hi\ envelope which surrounds the
Magellanic Clouds, and is thus 50 to 60 kpc distant.  The distance of
MS II must be similar; the distance of MS IV is less certain, but from
30 to 60 kpc in most models.  Thus these observations imply
a coronal density $n_{\rm H} \sim 10^{-4}~\cmcub$ at $\sim 50$~kpc
above the galactic plane, since the Stream clouds detected are at
$b =$~-70\deg\ to -80\deg.  The corona is presumably of equal or greater
extent in the plane of the Galaxy.

This implies that the Galaxy has a very large {\it baryonic}, gaseous
extent, much larger than previously known; studies of gas in the
Galactic halo have shown high-$z$ gas extending out of the disk,
but probe only to scale heights of several kpc (\eg Savage \& de Boer 1979,
Reynolds 1991, Danly 1992, Albert \etal 1994), while studies of the
X-ray background have suggested a gaseous corona (\eg Wang 1992),
but cannot give distance information.  This extended corona
is a significant phase of the interstellar medium of the Galaxy.
Assuming it is distributed smoothly, a rough lower limit
on its mass is $2 \times 10^9~\msun$, the mass of a sphere of
primordial gas with constant density $n_{\rm H} = 10^{-4}~\cmcub$
and radius 50 kpc.

The most immediate consequence of such an extensive gaseous corona is
to suggest that ram pressure stripping is significant in the origin
of the Magellanic Stream.  The density implied for the coronal
gas is very similar to that assumed in the model of Meurer \etal (1985),
for example; however, their best model requires an extremely high drag
coefficient $C_D$, implausible given the present \ha\ measurements;
the model of Moore \& Davis (1994) overcomes this difficulty, but
requires the additional presence of a extended ionized gaseous disk.

Secondly, this coronal gas is in rough pressure equilibrium with the
Stream clouds, if the clouds are at $T \sim 10^4~\degk$.
This pressure confinement is necessary to
stabilize the Stream clouds (Mirabel, Cohen \& Davies 1979).
It is pleasing to note that this echoes the cloud-confinement
argument which led Spitzer (1956) to propose the presence of hot gas
in the halo of the Galaxy.

The present result also provides evidence supporting the conjecture
of Bahcall \& Spitzer (1969) that normal galaxies have large gaseous
halos, which they invoked to account for
QSO absorption lines, presumed to arise in low column-density
neutral systems within the halos.  Recent work (Lanzetta \etal 1994)
suggests that many low-redshift absorption systems are indeed associated
with galaxies,
which requires normal galaxies to have a gaseous extent of
$\sim 160h^{-1}$~kpc.  This is compatible with the coronal density
at 50 kpc suggested in this paper (Mo 1994).

Additionally, it has long been known that a hot ionized corona
could contribute to the observed soft X-ray background (\eg Silk 1974,
McCammon \etal 1983).  The detection of an X-ray
shadow due to the Draco Nebula shows that there is soft X-ray emitting
gas outside the Local Bubble, but gives only a lower limit of
$\sim 600$ pc for the distance (Snowden \etal 1991, Burrows \& Mendenhall
1991).  These results also suggest that any corona may be patchy
or not smoothly distributed.
Wang (1992) concludes that a corona of density
$n \sim 10^{-4}~\cmcub$ at 50 kpc is consistent with the X-ray background.

In conclusion, we have detected \ha\ emission from several clouds of
the Magellanic Stream, of EM 0.5 -- 1 \cmsixpc.  The emission is
best explained by ram pressure heating of the Stream as it moves through
hot ionized coronal gas, of density $n_{\rm H} \sim 10^{-4}~\cmcub$
at a Galactic radius $R \sim 50$~kpc.  This implies that the
Galaxy has a very large baryonic, gaseous extent, in accord with
recent studies of low-redshift QSO absorption lines, and
supports ram-pressure models for the origins of the Magellanic
Stream.  Further observations and modeling of the Stream may
provide an unprecedented opportunity to probe gas in the outer Galactic
halo.

\acknowledgements
It is a pleasure to thank the CTIO staff for their customary excellent
support, particularly Luis Gonzalez for his assistance at the telescope.
We thank the referee, Knox Long, for an important correction regarding
shocks.  BW thanks Povilas Palunas and Karl Gebhardt for advice and
encouragement.  We are also grateful to Illimani, God of the Andes,
for propitious weather.  The RFP was developed with support from Rutgers
University and the NSF, under grant AST-8319344.

\clearpage


\begin{figure}
\caption[]{
(a) Spectrum from one 15-minute exposure on the MS II A
field.  The strong feature at 6563~\AA\ is geocoronal \ha; the smaller
feature at 6560~\AA\ is associated with the MS II cloud.  The size
of the symbols is approximately the size of the error bars.

(b) As (a) but for the corresponding sky field.

(c) The sky-subtracted spectrum; a constant continuum estimated
from regions away from the emission has been subtracted.  The error
bars are shown.}
\end{figure}

\begin{figure}
\caption[]{
Total spectra, with sky and continuum subtracted, for three
object fields located at leading edges of \hi\ concentrations in the Stream.
The error bars on individual points are
omitted for clarity.  The solid line is the LOWESS--smoothed fit to
the data.  The inverted triangles indicate the blueshift of \ha\
expected from the \hi\ velocity given by 21 cm observations.

(a) MS II, field A, on the leading edge of the MS II cloud.  The
fluctuations around 6563~\AA\ are residuals from the geocoronal \ha\ line.

(b) MS III, leading edge.

(c) MS IV C, leading edge.
}
\end{figure}

\begin{figure}
\caption[]{
Total spectra, with sky and continuum subtracted, for four fields on the
Stream, but not on cloud leading edges.

(a) MS II B.  Some points near 6563~\AA\ lie outside the range of the
graph, due to large residuals from the geocoronal \ha\ line.

(b-d) MS IV A, MS IV B, and MS IV D.}
\end{figure}

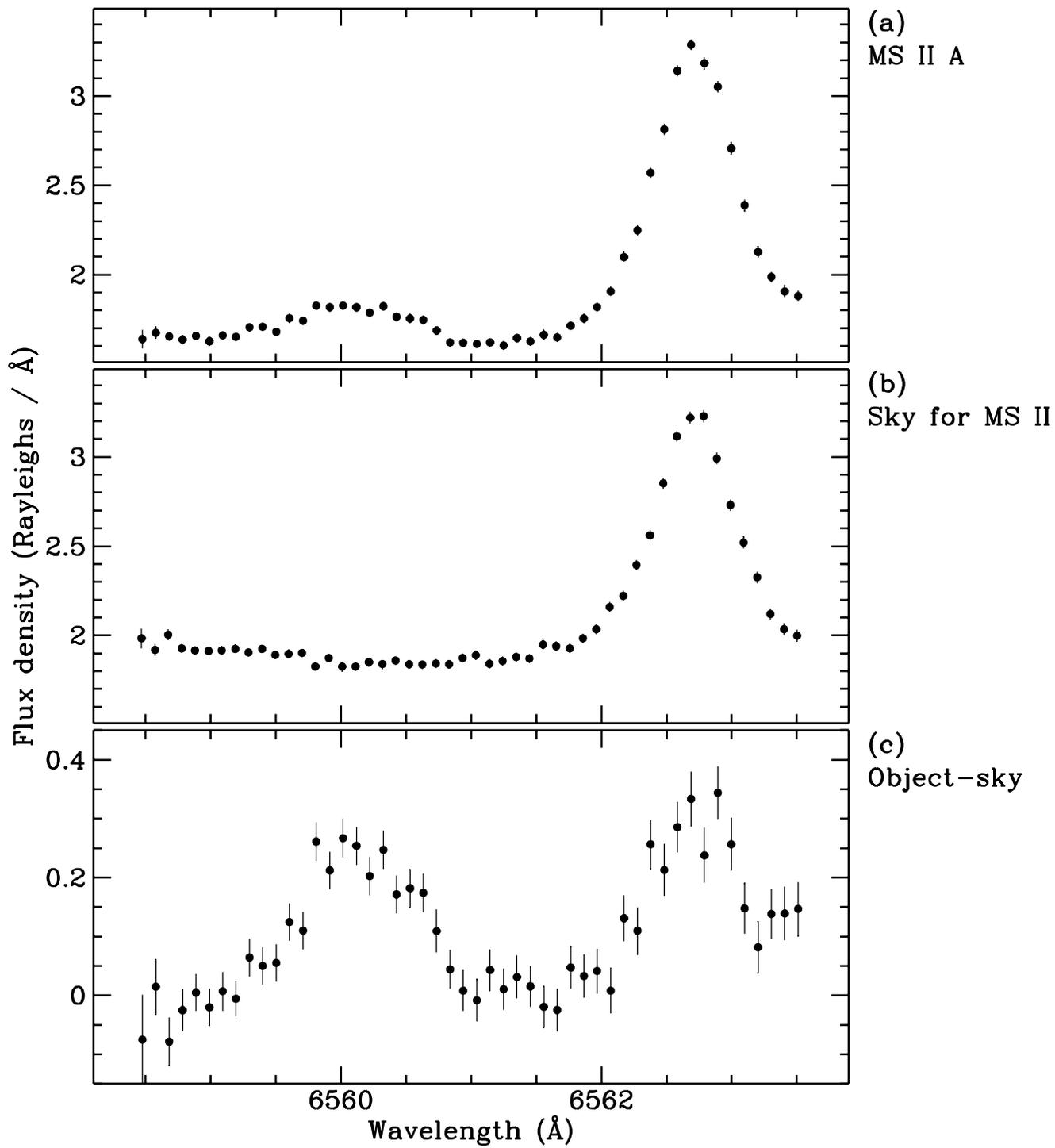
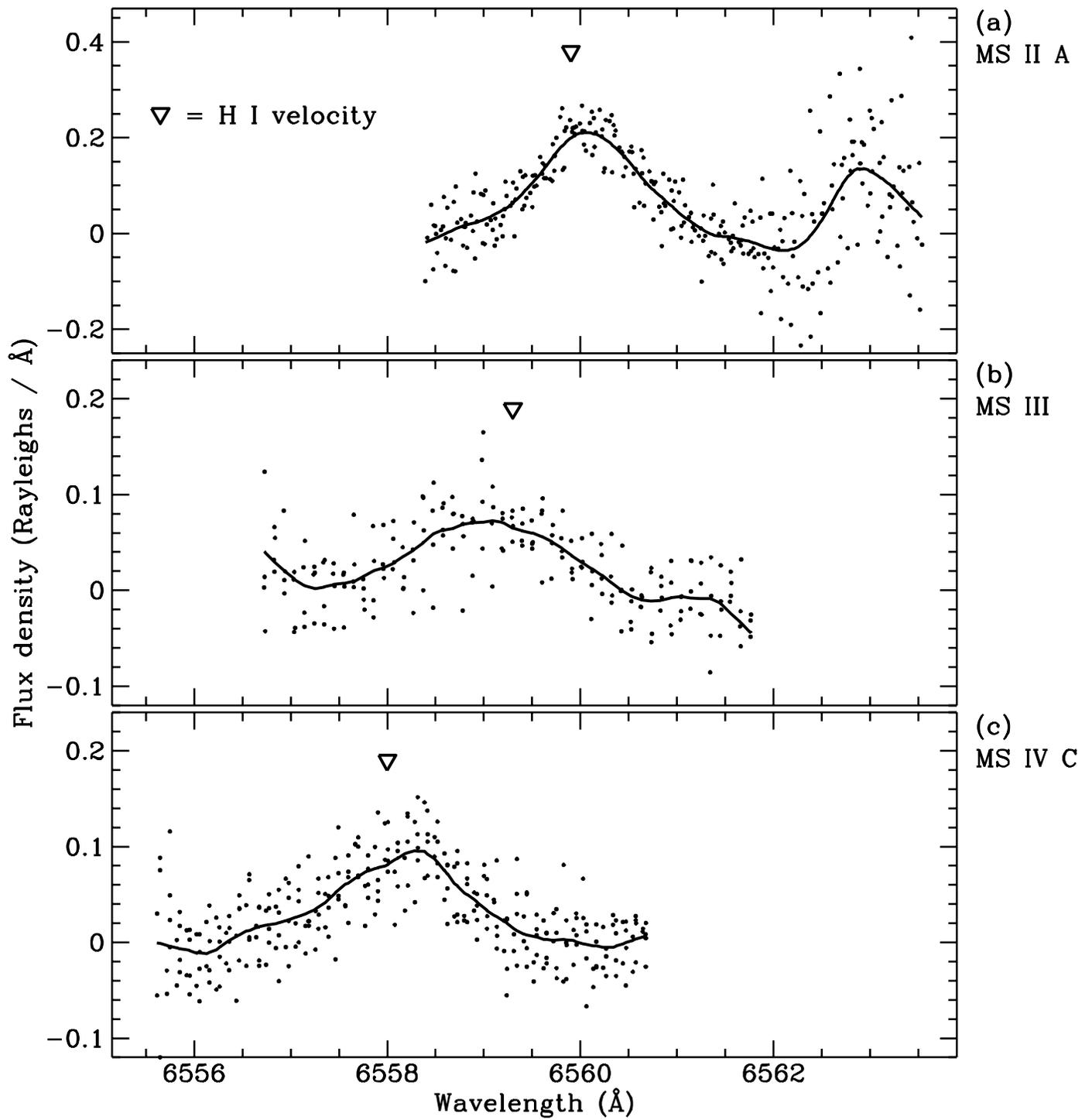
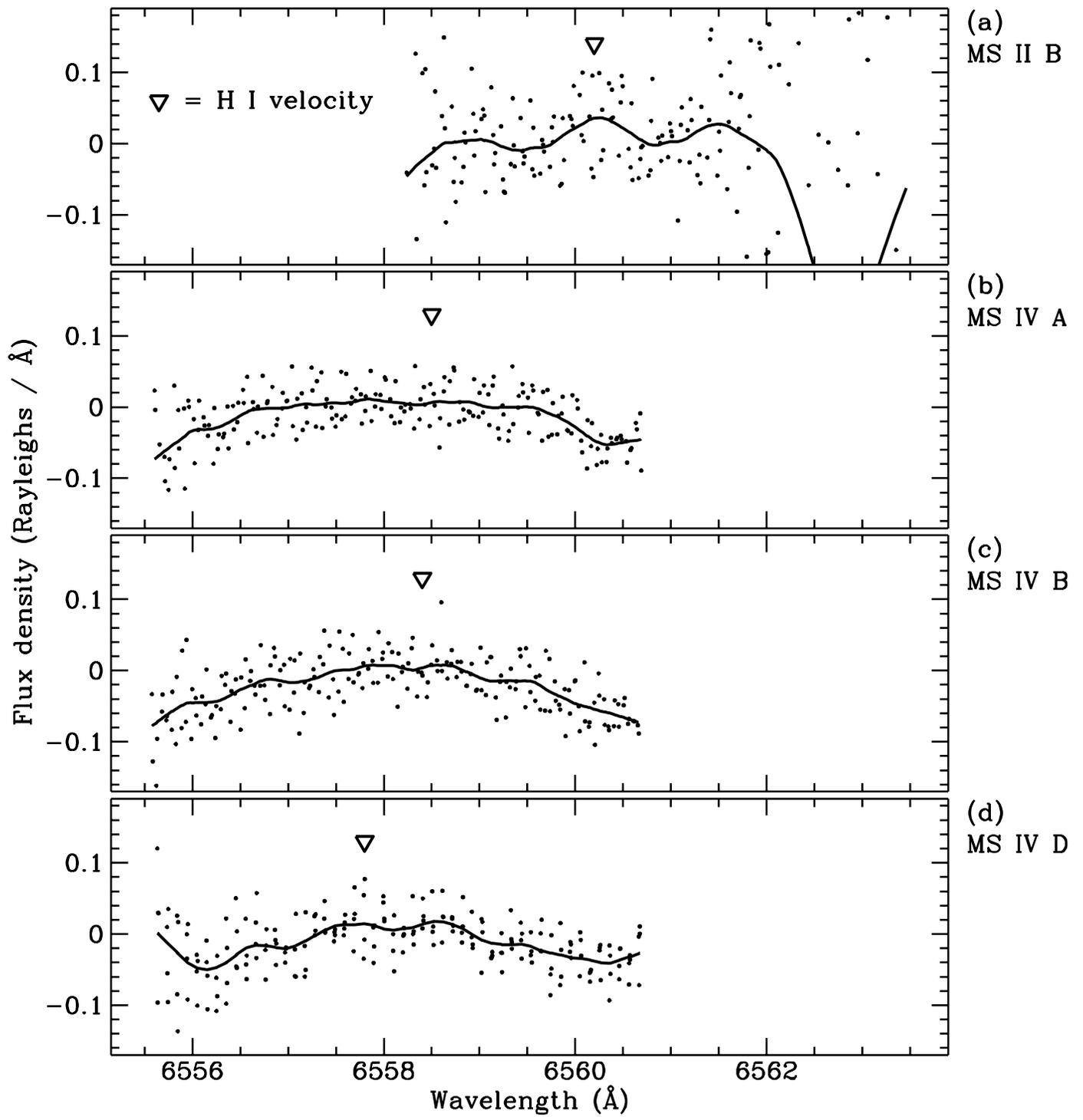
\end{document}